\documentclass{aastex}
\usepackage{spr-astr-addons}
\usepackage{url}
\usepackage{graphicx}
\usepackage{tabularx}
\usepackage{longtable}

\RequirePackage{color}

\newcommand{\emaila}{burak.ulas@comu.edu.tr}

\begin{document}

\title{Discovery of pulsating components in eclipsing binary systems through the TESS light curves: The cases of CPD-30~740, HD~97329, V1637~Ori and TYC~683-640-1}

\slugcomment{Not to appear in Nonlearned J., 45.}
\shorttitle{Discovery of pulsating components}
\shortauthors{Ula\c{s} et al.}

\author{B.~Ula\c{s}\altaffilmark{1,2}} 
\email{\emaila}
\author{C.~Ulusoy\altaffilmark{3}}
\author{N.~Erkan\altaffilmark{4,2}}
\author{M.~Madiba\altaffilmark{3}}
\author{M.~Matsete\altaffilmark{3}}

\altaffiltext{1}{Department of Space Sciences and Technologies, Faculty of Arts and Sciences, \c{C}anakkale Onsekiz Mart University, Terzio\v{g}lu Campus, TR~17100, \c{C}anakkale, Turkey.}
\altaffiltext{2}{Astrophysics Research Centre and Observatory, \c{C}anakkale Onsekiz Mart University, Terzio\v{g}lu Campus, TR~17100, \c{C}anakkale, Turkey.}
\altaffiltext{3}{Department of Physics and Astronomy, Botswana International University of Science and Technology (BIUST), Plot~10017, Private Bag 16, Palapye, Botswana.}
\altaffiltext{4}{Department of Physics, Faculty of Arts and Sciences, \c{C}anakkale Onsekiz Mart University, Terzio\v{g}lu Campus, TR~17100, \c{C}anakkale, Turkey.}

\begin{abstract}
We present the first evidence for the pulsations of primary components of four eclipsing binary systems. The TESS light curves of the targets are analyzed, the light and the absolute parameters are derived. Fourier analyses are applied to the residual data subtracted from the binary models to reveal the pulsational characteristics of the primaries. The components of the systems are compared to well-known binary systems of the same morphological type. The types of pulsational classes for the primaries are also discussed. Results show that  CPD-30~740 and V1637~Ori are oscillating eclipsing Algol type binary stars, while HD~97329 and TYC~683-640-1 are found to be detached systems with pulsating primary components. We conclude that the primary components of the systems are pulsating stars.
\end{abstract}



\section{Introduction}
Eclipsing binary systems are essential to obtain precise measurements of the fundamental astrophysical parameters of stars. They consist of components with a wide range of spectral types of pulsating stars in the HR diagram \citep{koo16,lia17}. Some of the components can be found in the group of variables within the instability strips (e.g. $\delta$~Sct and $\beta$~Cep stars). The existence of such a variable component in an eclipsing binary system represents as a unique opportunity to minimize physical constraints by successfully applying asteroseismology to high-precision space-based data \citep[e.g.][]{sou11,lee18,ant19,lee20}. Eclipsing binary systems with $\delta$~Sct pulsators furthermore advantageous to derive more information for investigating the complex interaction between the sensitivity of stellar acoustic frequency spectra, rotation, and binary evolution \citep{mor13}. 

$\delta$~Sct stars are generally defined as Population I stars with a spectral type between A0-F5 \citep[temperatures correspond between 7000 and 8500~K respectively;][]{cha13}. Their location is close to an extension of the classical instability strip in the H-R diagram. Their pulsations are driven by $\kappa$-mechanism in radial and non-radial pressure and gravity modes, or both \citep{gri10,uyt11} with periods in the range between about 0.008-0.42~days \citep{cat15}.

Eclipsing binary systems with pulsating components of $\delta$~Sct type can be classified as detached \citep[EA/DSCT,][]{pop80,and91} and semi-detached systems \citep[oEA -- oscillating Eclipsing Algol System;][]{mkr02,mkr04} according to binary configuration. The oEA stars comprise eclipsing Algol stars with where the primary mass accreting shows $\delta$~Sct type pulsational behavior. Thus, these stars are exceptional objects for investigating short-term dynamic stellar evolution during mass transfer episodes, presumably due to the secondary component’s magnetic activity cycle, and they are located inside the instability strip. \cite{mkr18} showed for the first time that mass transfer and accretion influence the amplitudes and frequencies of the non-radial pulsation modes for the oEA type star RZ~Cas.  

Recent space missions such as CoRoT \citep{bag06}, {\it{Kepler}}/K2  \citep{bor10,gil10,how14} and TESS \citep{ric15} have dramatically improved our understanding of various phenomena of the stellar variability with a large number of newly detected eclipsing binary systems with pulsating components. The systems presented in this study have not been studied in depth in the literature, in particular in terms of their binary and pulsating properties. 

CPD-30~740 (TIC~737546) is in the catalog of \cite{chr08} who listed $J$, $H$, $K$, and $I$ magnitudes of the star. The $B$, $V$ magnitudes, and effective temperature for CPD-30~740 cataloged by \citep{mun14}. The $uvby\beta$ magnitudes from HD~97329 (TIC~81702112) were cataloged by \cite{hau98} and \cite{pau15}. In addition, \cite{ren09} included the system in their Ap, HgMn, and Am stars catalog. A more detailed study on the system was carried out by \cite{koz16} who performed spectroscopic observations of the binary and measured the radial velocities. They analyzed the ASAS light curve in combination with their radial velocity data and derived binary characteristics of the system. The light variation of V1637~Ori (TIC~244208023) was originally reported by \cite{hof44}. \cite{gar00} identified the target as an eclipsing binary based on their observed light curve. V1637~Ori also listed in the eclipsing binary catalog by \cite{avv13} and Catalina surveys of \cite{dra14}. \cite{can18} concluded that the probability of being an open cluster member for TYC~683-640-1 (TIC~450089997) was zero.  All of our targets have also been published on Gaia DR1 and DR2 \citep{gai16,gai18}.

In this study, we present the comprehensive analyses of the pulsational properties of four eclipsing binary systems in the TESS field. The paper is structured as follows: we present the description of the structure and properties of observational data in the next section. The third section focuses on the analysis of light curves through modeling of binary configurations. The frequency analyses and pulsational characteristics are presented in \ref{s:pulse}. We therefore discuss our results and give the concluding remarks in the last section.

\section{Light Curve Data}\label{s:data}

\begin{figure*}
\centering
\includegraphics{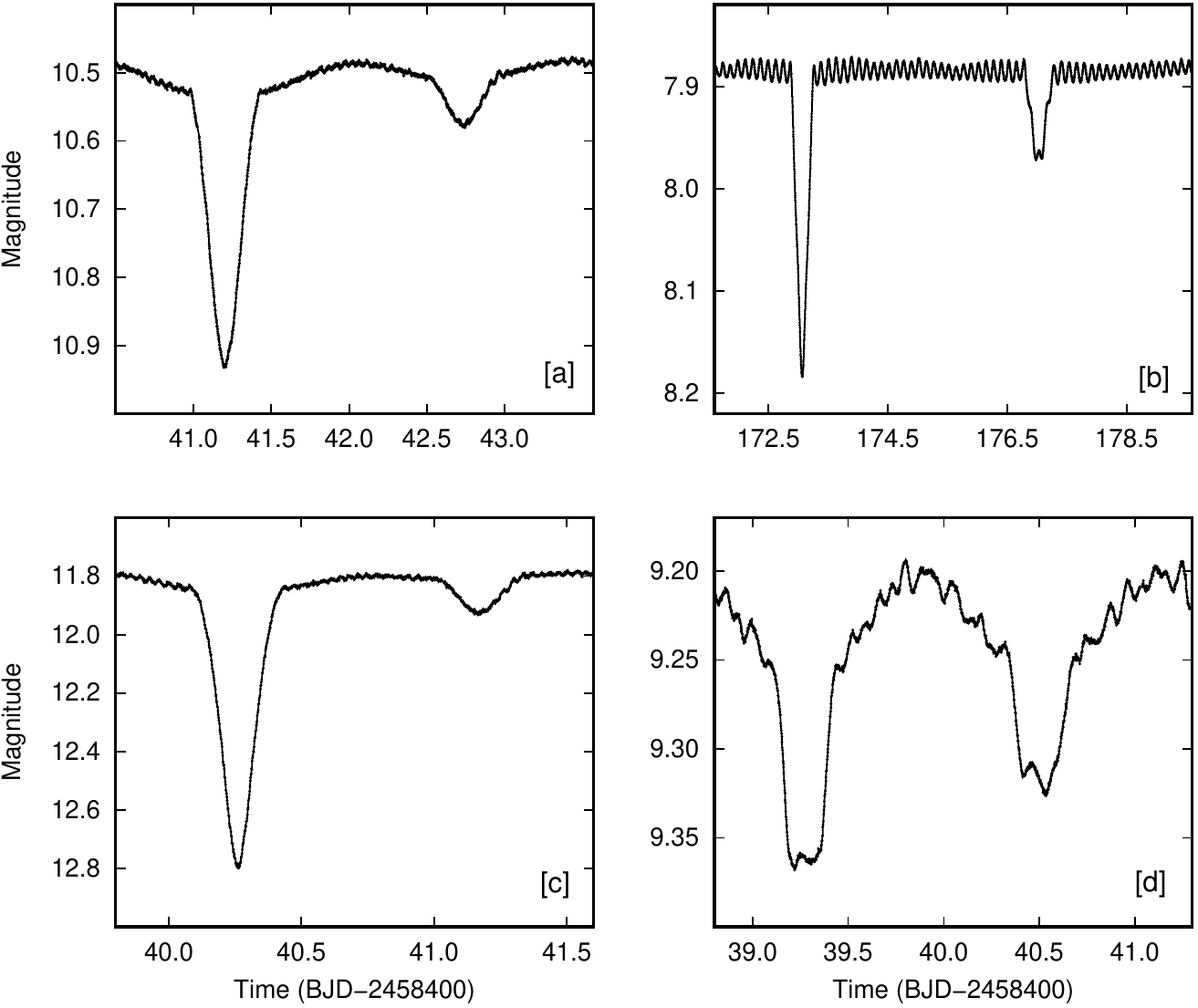}
\caption{One-period-long light curves of CPD-30~740 (a), HD~97329 (b), V1637~Ori (c), and TYC~683-640-1 (d). Magnitudes are calculated by using the TESS magnitudes given by \citep{sta19}}
\label{fig1}
\end{figure*}

We used 2-minute cadence photometric time-series data provided by TESS which are publicly available through {\sc MAST} (Mikulski Archive of Space Telescope Portal)\footnote{mast.stsci.edu/portal/Mashup/Clients/Mast/Portal.html}. Time and PDC\_SAP (Pre-search Data Conditioning Simple Aperture Photometry) flux values were extracted from the data and the flux values ($F_i$) were converted to magnitudes ($m_i$) by using $m_{i} = -2.5 \log F_{i}$ formula to yield the light curves. Fig.~1 shows the light curves in their one-orbital-period-long time duration to ensure clear visibility of the pulsational effects on their light.  
The shapes of the light curves allow us to assume a superposition of multiperiodic pulsational light variation on a typical eclipsing binary light curve (Fig.~\ref{fig1}) as seen in many previous systems investigated in the literature \citep{kim03}. The duration of primary and secondary minima are about 11 and 9 hours for CPD-30~740, 9.5, and 12 hours for HD~97329, 8, and 7.5 hours for V1637~Ori, 8.5, and 9.0 hours for TYC~683-640-1. The data files cover 17673 (26 days), 15602 (24.7 days), 17664 (26 days), and 17666 (26 days) data points for CPD-30~740, HD~97329, V1637~Ori, and TYC~683-640-1, respectively, excluding {\tt{NULL}} data.

The pulsating components are primaries, which are eclipsed during primary minima, since oscillations are dominant in the secondary minima.

\section{Binary Modelling} \label{s:binary}
\begin{figure*}
\centering
\includegraphics{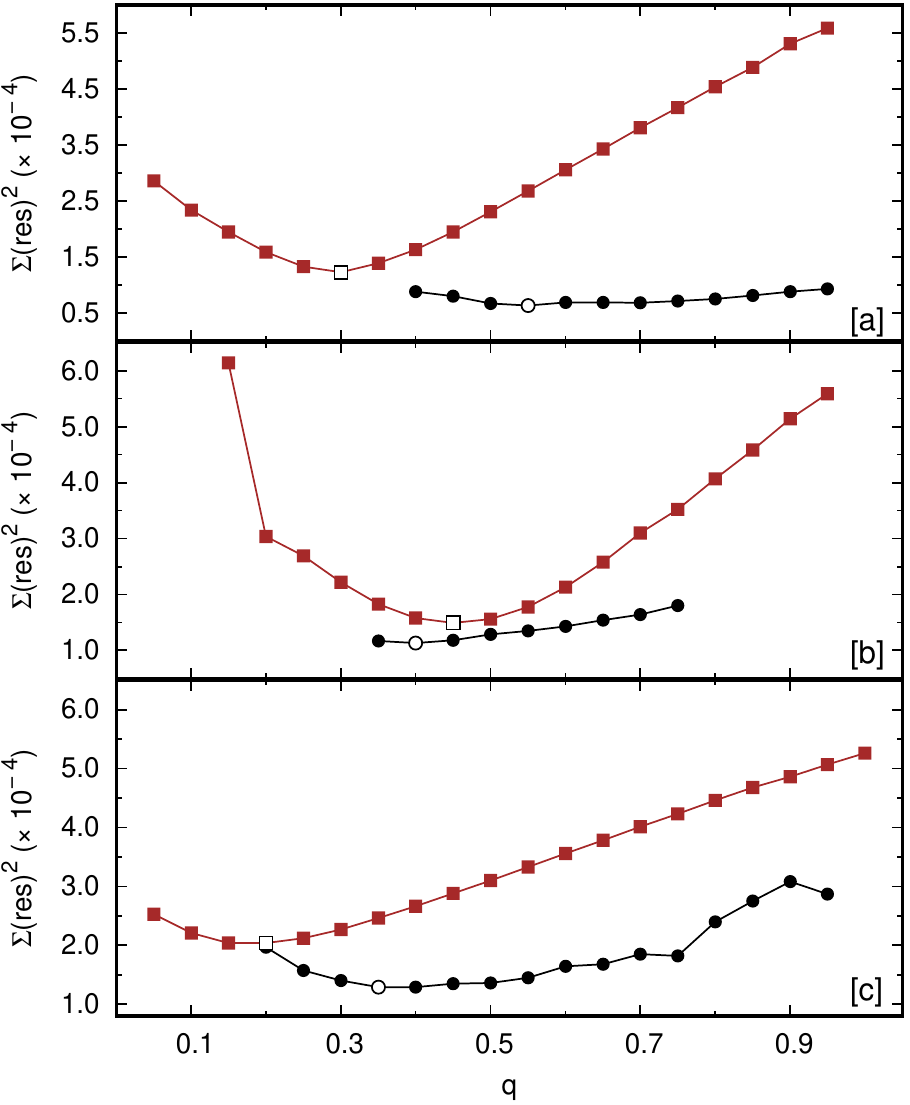}
\caption{Results of the $q$-search for CPD-30~740 (a), V1637~Ori (b), and TYC~683-640-1 (c) with detached (circles) and semidetached (squares, brown in colored version) configuration assumptions. The values having the minimum $\sum \left(res \right)^{2}$ are represented with open signs}
\label{figq}
\end{figure*}

Our analyses represent the first light curve analyses for the systems CPD-30~740, V1637~Ori, and TYC~683-640-1 in the literature. The lack of reliable parameters enables us to start the solution by applying a $q$-search to normalized light curves covering 1000 data points using the 2015 version of Wilson-Devinney code \citep{wil71,wil20}. The $q$-search proceeded in two different morphological configurations; the detached binary and semidetached binary with a secondary star fills its Roche lobe. The results are given in Fig.~\ref{figq}.

Following the derivation of initial mass ratios, the analyses of the light curves were made using the {\tt PHOEBE} \citep{pri05} software, which employs the Wilson-Devinney method  \citep{wil71} to achieve the stellar parameters from the light curve (and radial velocities curve for HD~97329) data. 

We applied analyses with two different assumptions that the systems are in detached and conventional semidetached configurations. More precise results were achieved in models with detached configuration for HD~97329 and TYC~683-640-1. It is found out that the semi-detached configuration is the most proper assumption for CPD-30~740 and V1637~Ori considering the factors given in relevant subsections. The following parameters are set as adjustable during the analyses; The inclination $i$, the temperature of the secondary component $T_2$, mass ratio $q$, the surface potential value of primary $\Omega_1$, luminosity of the primary component $L_1$. The surface potential value of secondary $\Omega_2$ was also set as free parameter during the solutions with the detached configuration assumptions. Keeping in mind that the granulation boundary for main sequence stars located at about F0 spectral type according to \cite{gra89}, the albedos ($A_{1}$ and $A_{2}$) are calculated from \cite{ruc69} and the gravity darkening ($g_{1}$ and $g_{2}$) coefficients for the systems was adopted from \cite{zei24} and \cite{luc67}. Logarithmic limb darkening coefficients ($x_{1}$ and $x_{2}$), on the other hand, were calculated from \cite{cla17} according to the initial temperatures of the components.

The orbital periods of the systems adopted in the light curve solutions were derived by period analyses. Since there is no enough number of times of minimum light in the literature we calculated minimum times from TESS light curves using the method of \cite{kwe56} and combined them with few times of minimums in the literature (Table~\ref{tmin}). The period analyses were conducted by assuming that the variation is linear and changes in time of minimum light ($\Delta T_0$) and orbital period ($\Delta P$) were calculated using the equation $O-C=\Delta T_{0}+\Delta P\cdot E$, where $E$ is the cycle number. Results are listed for each system in the first two lines of Table~\ref{tlc} with the light parameters.

\begin{table*}
\centering
\caption{Times of minima derived from the TESS light curves and collected from the literature}
\label{tmin}
\begin{tabular}{llccllcc}
\hline
Star & \multicolumn{1}{c}{BJD}     & Type  & Ref. & Star & \multicolumn{1}{c}{BJD}     & Type  & Ref.	\\
\hline
CPD-30~740&2458439.66936(11)&sec.&1&V1637~Ori&2458446.6378(4)&sec.&1\\
&2458441.20167(2)&pri.&1&&2458447.55169(6)&pri.&1\\
&2458442.7363(1)&sec.&1&&2458448.4599(4)&sec.&1\\
&2458444.26968(2)&pri.&1&&2458452.1057(4)&sec.&1\\
&2458445.8026(1)&sec.&1&&2458453.01994(5)&pri.&1\\
&2458447.33743(2)&pri.&1&&2458453.9295(4)&sec.&1\\
&2458451.9405(1)&sec.&1&&2458454.84245(5)&pri.&1\\
&2458453.47277(2)&pri.&1&&2458455.7515(4)&sec.&1\\
&2458455.0080(1)&sec.&1&&2458456.66513(5)&pri.&1\\
&2458456.54055(2)&pri.&1&&2458457.5739(5)&sec.&1\\
&2458458.0761(1)&sec.&1&&2458458.48808(5)&pri.&1\\
&2458459.60799(2)&pri.&1&&2458459.3976(4)&sec.&1\\
&2458461.1438(1)&sec.&1&&2458460.31080(5)&pri.&1\\
&2458462.67572(2)&pri.&1&&2458461.2199(4)&sec.&1\\
HD~97329&2451873.96(1)&pri.&2&&2458462.13355(5)&pri.&1\\
&2458573.07177(3)&pri.&1&TYC~683-640-1&2458439.27308(6)&pri.&1\\
&2458577.0311(1)&sec.&1&&2458440.5019(2)&sec.&1\\
&2458588.54300(3)&pri.&1&&2458441.73789(7)&pri.&1\\
&2458592.5006(1)&sec.&1&&2458442.9677(2)&sec.&1\\
V1637~Ori&2450775.5244&pri.&3&&2458444.19546(7)&pri.&1\\
&2451428.0647&pri.&3&&2458445.4410(3)&sec.&1\\
&2455862.8902&pri.&3&&2458446.65986(7)&pri.&1\\
&2458438.43750(5)&pri.&1&&2458447.8896(2)&sec.&1\\
&2458439.3457(4)&sec.&1&&2458454.04534(6)&pri.&1\\
&2458440.26046(5)&pri.&1&&2458452.8154(2)&sec.&1\\
&2458441.1692(4)&sec.&1&&2458456.50461(7)&pri.&1\\
&2458442.08330(5)&pri.&1&&2458455.2716(2)&sec.&1\\
&2458442.9914(4)&sec.&1&&2458458.96749(6)&pri.&1\\
&2458443.90545(5)&pri.&1&&2458457.7404(2)&sec.&1\\
&2458444.8137(4)&sec.&1&&2458461.43105(7)&pri.&1\\
&2458445.72884(5)&pri.&1&&2458460.1975(2)&sec.&1\\
\hline
\multicolumn{8}{l}{References: 1. This study, 2. \cite{koz16}, 3. \cite{pas06}} \\
\end{tabular}
\end{table*}

\subsection{CPD-30~740}\label{cpd}
\begin{figure}
\centering
\includegraphics[scale=0.85]{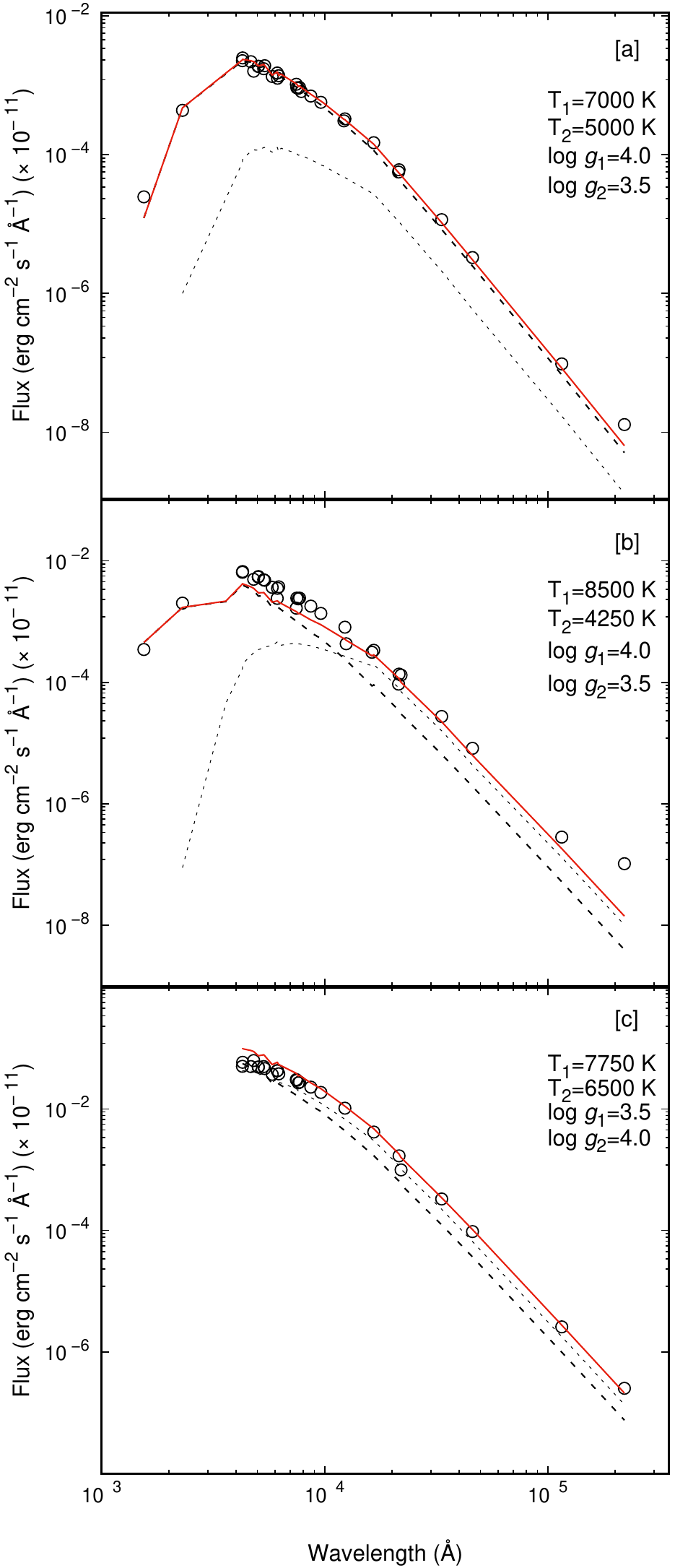}
\caption{Results and fits from the SED analyses for CPD-30~740 (a), V1637~Ori (b), and TYC~683-640-1 (c). Open circles indicate the photometric flux values from the VizieR database \citep{och00}. The best model fits for the primaries plotted with a dashed line, while the dotted lines symbolize the fits for the secondary components. The solid (red in colored version) lines refer to the total model fits resulted with the effective temperatures and log$g$ values of the components given on the upper right of the figures. Note that the axes are in logarithmic scale}
\label{fsed}
\end{figure}

Light curve analysis of the system was conducted with the effective temperatures for the components that derived by constructing a spectral energy distribution (SED) through Virtual Observatory SED Analyzer \citep[VOSA,][]{bay08} for the available photometric data in the VizieR database \citep{och00}. The photometric data were fit with the assumption of two components (binary fit) by using a parameter-grid search in order to obtain the optimized Kurucz atmosphere model \citep{kur79}. During the SED analysis, the minimum and maximum values for effective temperature were set 7000 and 9000~K for primary, 4000 and 6000~K for secondary. The log$g$ values, on the other hand, were assumed to be between 3.5-4.0 and 3.0-3.5 for primary and secondary components, respectively. The extinction, A$_{\nu}$, was set to 0.0287$^m$ which is calculated using the Galactic Dust Reddening and Extinction interface of NASA/IPAC Infrared Science Archive\footnote{https://irsa.ipac.caltech.edu/applications/DUST/}. The analysis resulted in the values of T$_1$=7000$\pm$125, T$_2$=5000$\pm$125,  log$g_1$=4.0$\pm$0.25 and log$g_2$=3.5$\pm$0.25. Vgf$_b$, the modified $\chi^2$ of the solution, which is calculated by compelling the error in observed flux to be larger than 0.1 times of the corresponding flux value, was found to be 1.7. Vgf$_b < 10-15$ corresponds to a good fit \citep{bay08}, therefore, these values were adopted for the components during the light curve solution. The model fit of the SED analysis is plotted in Fig.~\ref{fsed}. 

The orbital period value in the light curve solution adopted from our period analysis, which bases the initial value of 3.067739 days given by All Sky Automated Survey database \citep{poj02}. Although the $q$--search results provides that the system is a detached binary, the similarity of the system's light curve to a typical semi-detached Algol binary having A-type primary and cool low mass secondary component lead us to investigate advanced clues to decide the certain morphological class of the system. Therefore, we analyzed the light curve of the system with the assumption of both semi-detached (with Roche lobe filling secondary) and detached configurations to compare the results and determine the geometrical configuration.

\begin{figure}
\centering
\includegraphics[scale=0.78]{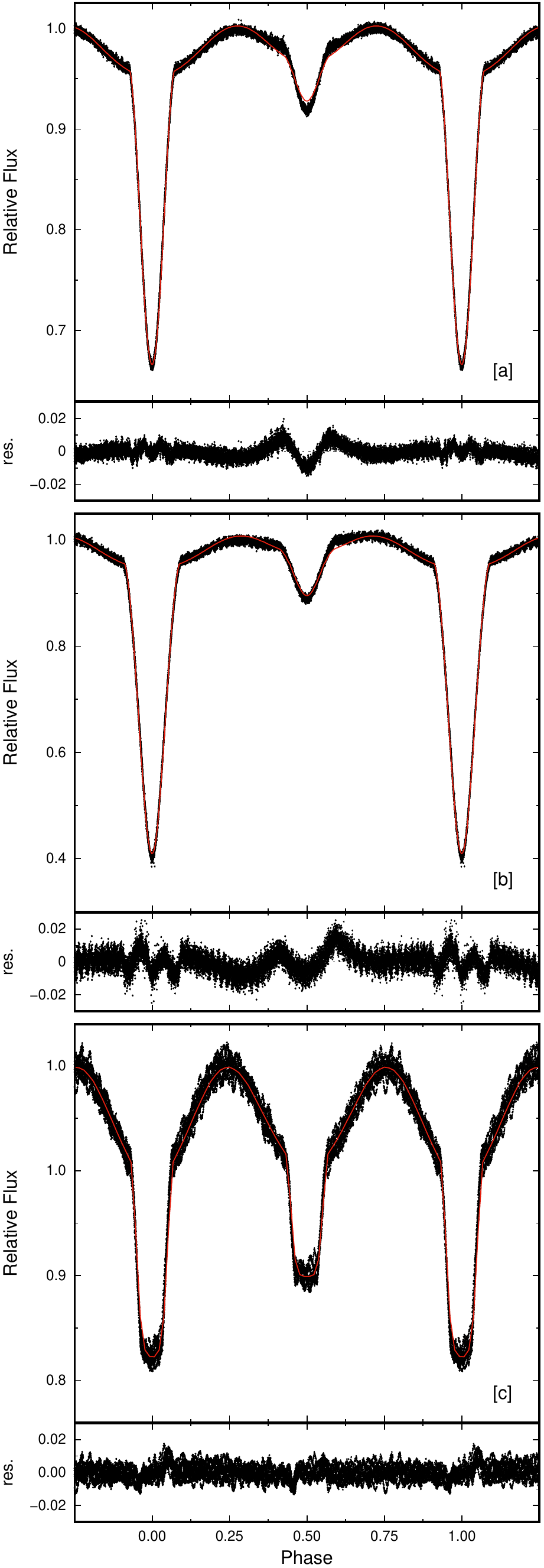}
\caption{Synthetic light curves (lines, red in colored version) calculated from the analyses are shown with the observations for CPD-30~740 (a), V1637~Ori (b), and TYC~683-640-1 (c). Residuals from the solution are also shown at the bottom panels of related light curves}
\label{flcs}
\end{figure}

The initial mass ratio was 0.3 and 0.55 for semi-detached and detached models, the results of our $q$--search as mentioned previously. The albedo for the secondary component were also set free during the analyses to achieve the better fit around the secondary minimum. The observations are compared to resulting synthetic light curve for semi-detached assumption in Fig.~\ref{flcs} and calculated light parameters for both geometrical configuration are tabulated in Table~\ref{tlc}. Our analyses correspond to the secondary component to be a K type star, which is also typical spectral type for the secondaries of Algol type binaries. Indeed, even in the detached solution, the secondary component of the system found to be filled 95\% of its Roche lobe following the equation $1-$ \scalebox{1.2}{$\frac{\Omega_{2} - \Omega_{cr}}{\Omega_{cr}}$}, where $\Omega_{2}$ and $\Omega_{cr}(=2.969)$ refer to the quantities in Table~\ref{tlc}. Combining the results of the analyses with the argument on the shape of the light curve mentioned in the previous paragraph, we indicate that the system is most likely a semi-detached system with A-F type primary and a low mass giant or sub-giant secondary. Thus, our further calculations were made by considering the results of semi-detached binary model analysis.

\begin{table*}
\centering
\small
\caption{Results of the light curve analyses. Orbital period values are derived by period analyses mentioned in the text. 2450000 is subtracted from the times of minimum lights. $r_1$ and $r_2$ refer to the fractional radii of the components. Uncertainties in $T_1$ of CPD-30~740, V1637~Ori, and TYC~683-640-1 are given as obtained in SED analyses. SD and D stands for the results for semi-detached and detached configuration, respectively. $\Omega_{cr}$ refer to the critical potential value
of the first Lagrangian point. The standard deviations, 3$\sigma$ for the last digits of light parameters are given in parentheses}
\label{tlc}
\begin{tabular}{lccccc}
\hline
Parameter & CPD-30~740 [SD]       & CPD-30~740 [D]    & HD~97329 	  & V1637~Ori		& TYC~683-640-1	\\                      
\hline                                                                                                                                               
T$_0$ (BJD)         & 8441.20167(2)        & 8441.20167(2)     & 8573.07177(3) & 8438.43750(5)     & 8441.73789(7) \\                         
$P$~(days)                 & 3.06774(4)           & 3.06774(4) & 7.73576(2)    & 1.82269(8)        & 2.46170(6) \\                            
$i$ ${({^\circ})}$         & 73.31(2)   	  & 72.45(1)   	& 82.88(3)          & 77.61(2)     		& 82.41(2)     	\\                           
$q$                        & 0.311(1)   	  & 0.549(1)   	& 1.1(1) 	    	& 0.555(2)   		& 0.410(1)   		\\           
$T_1$ (K)                  & 7000(125)            & 7000(125)        & 6855      	& 8500(125)      	& 7750(125)      		\\   
$T_2$ (K)                  & 4611(42) 	          & 4842(27) 		& 5499(27)  	& 5296(26) 		    & 6808(24) 		\\           
$\Omega _{1}$              & 4.696(9)  	          & 5.051(6)  	& 11.38(3)      & 4.874(9)   		    & 3.427(3)   		\\                   
$\Omega _{2}$              & $\Omega_{cr}$=2.490  & 3.112(2)  	& 10.80(3)  	& $\Omega_{cr}$=2.980    & 5.297(9)    		\\                   
$r_1$                      & 0.229(3)  	          & 0.223(2)  	& 0.097(8)  	& 0.233(4) 	    	& 0.337(2) 		\\                   
$r_2$                      & 0.281(1)  	          & 0.298(3)  	& 0.111(8)  	& 0.327(1) 		    & 0.102(5) 		\\                   
$\frac{L_1}{L_1 +L_2}$     & 0.838(2) 	          & 0.768(1) & 0.671(1)         & 0.802(3) 	    	& 0.952(1) 		\\           
$A_{1},~A_{2}$ 		   & 1.0, 0.77 	          & 1.0, 0.77 		& 0.5, 0.5 	    &1.0, 0.7   		& 1.0, 0.5 		\\   
$g_{1},~g_{2}$ 		   & 1.0, 0.32 	          & 1.0, 0.32 	& 0.32, 0.32 	    & 1.0, 0.32 		    & 1.0, 0.32 		\\                   
$x_{1},~x_{2}$ 		   & 0.548, 0.670 	  & 0.548, 0.670 	& 0.554, 0.598      & 0.493, 0.706 		& 0.522, 0.547 	\\          
\hline
\end{tabular}
\end{table*}

\subsection{HD~97329}
\begin{figure}
\centering
\includegraphics[scale=0.77]{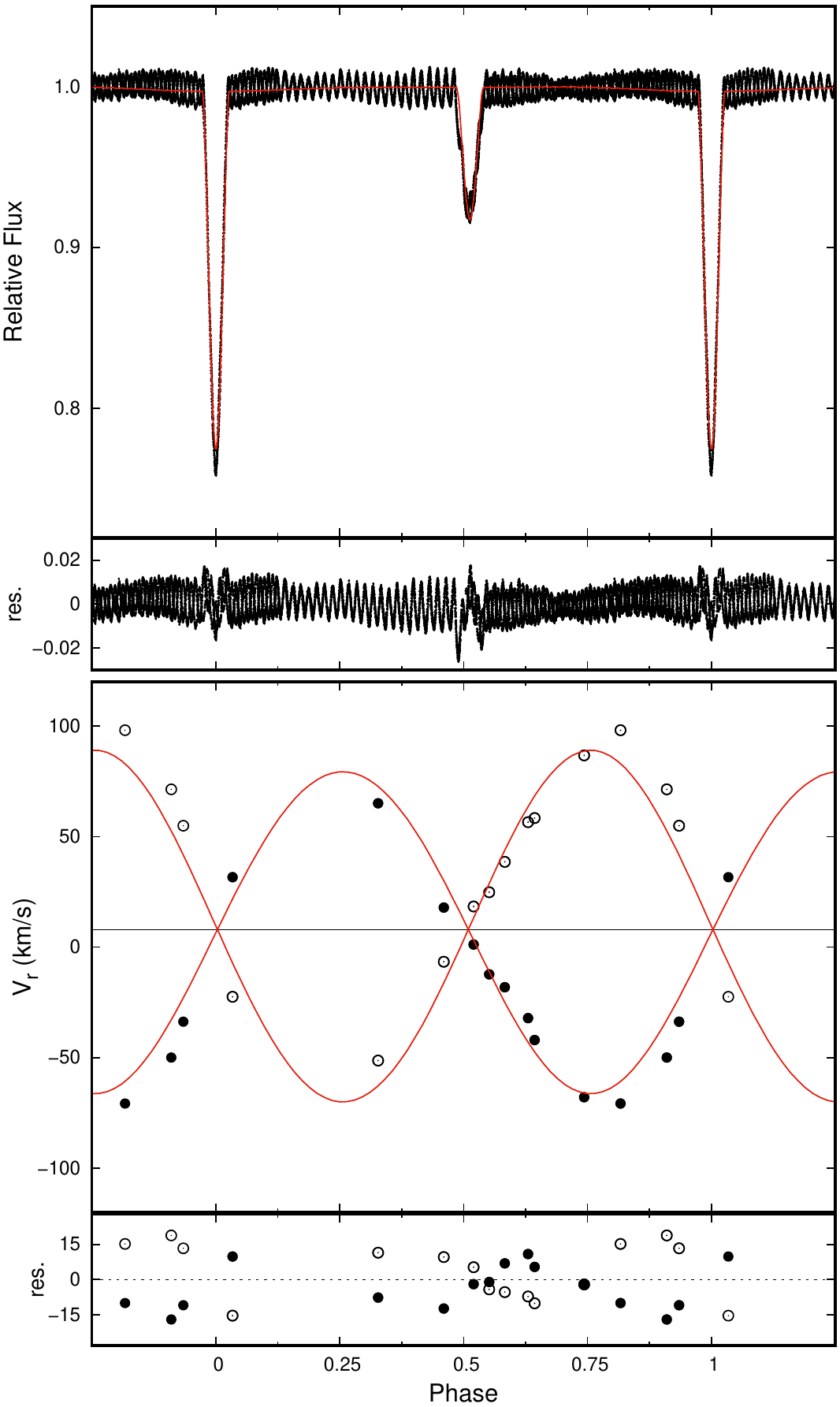}
\caption{Synthetic (line, red in colored version) and observational light curves of HD~97329 (upper panel). Calculated (line, red) and measured radial velocities are compared in the lower panel. The residuals are shown below the related curves. Radial velocities data are taken from \cite{koz16}}
\label{lchd}
\end{figure}

During the analysis, we adopted all initial parameters from \cite{koz16} which is the only detailed study on the binary parameters in the literature. The authors also obtained and listed the radial velocities measurements for the system. Therefore, we solved the TESS light curve and radial velocities curves simultaneously. In addition to the parameters mentioned at the beginning of Sec.~\ref{s:binary}, we also set the two other parameters as adjustable during the solution; eccentricity $e$, the velocity of the center of mass $V_0$. These two parameters were derived as $e$=0.0198$\pm$0.0001 and $V_0$=7.9$\pm$2.8~kms$^{-1}$. It should be noted that the hotter component is found to be slightly less massive than the cooler one. The results are listed in Table~\ref{tlc} and the light and radial velocities curves plotted in Fig.~\ref{lchd}.

\subsection{V1637~Ori }
Since there is no reliable value in the literature, we first derived effective temperatures for the components by analyzing the SEDs, the same method used for CPD-30~740 (See~\ref{cpd}). The analysis was applied with an initial effective temperature interval of 6000-9000~K for the primary and 4000-5500~K for the secondary component. Besides, the log$g$ intervals were set to 4.0-4.5 for primary and 3.0-3.5 for secondary. The extinction was 0.1083$^m$. The results are T$_1$=8500$\pm$125~K, T$_2$=4250$\pm$125~K, log$g_1$=4.0$\pm$0.25 and log$g_2$=3.5$\pm$0.25. Vgf$_b$ of the solution was 12.3. The fits of the SED analysis are compared to photometric data in Fig.~\ref{fsed}. 

We adopted the orbital period value of 1.82269 days from our period analysis, based on the value of 1.8226900101 days given by \cite{avv13}. Our $q$--search is not reliable in regard to decide the geometrical configuration of the binary, since it resulted in almost the same $\sum \left(res \right)^{2}$ values for semi-detached and detached assumptions. We decided to conduct the light curve solution with the semi-detached configuration based on the similarity of the light curve to a typical semi-detached Algol type binary. Similar to CPD-30~740, SED analysis of the star support the idea of being an Algol type binary by corresponding that the secondary component is potentially a K type giant or sub-giant. Moreover, a trial solution with detached configuration indicated that the secondary component filled 98\% of its Roche lobe. The initial mass ratio value of the analysis was 0.45 following our $q$--search. Setting the albedo for the secondary component free during the first steps of the analysis induced a better fit to observations in the neighborhood of 0.5 phase. The resulting parameters derived from the light curve analysis are given in Table~\ref{tlc} and the calculated light curve is plotted with the observations in Fig.~\ref{flcs}.

\subsection{TYC~683-640-1}
The system is not investigated in the literature so far. Therefore, determination for some initial parameters must be done to reach the realistic results in the light curve analysis. Hence, the effective temperature for the primary component was assumed by applying a SED analysis, as we did two of our targets in the study. The initial interval for the effective temperatures and log$g$ values were 7000-9000~K and 3.5-4.5 for primary and 6000-9000~K and 3.5-4.5 for secondary component. The visual extinction, on the other hand, was set to 0.4202$^m$, the value calculated from NASA/IPAC Infrared Science Archive. The results of the best fit (Fig.~\ref{fsed}) was T$_1$=7750$\pm$125~K, T$_2$=6500$\pm$125~K and log$g_1$=3.5$\pm$0.25, log$g_2$=4.0$\pm$0.25. Vgf$_b$ of the solution was found to be 0.22.

The orbital period of the binary is taken 2.46170 days, the result from the period analysis, during the solution. The initial mass ratio for the analysis was taken as the resulting value of the $q$-search, 0.35. Results are listed in Table~\ref{tlc} and synthetic and observational light curves are plotted Fig.~\ref{flcs}.

\section{Frequency Analyses} \label{s:pulse}

\begin{figure*}
\centering
\includegraphics[width=\textwidth]{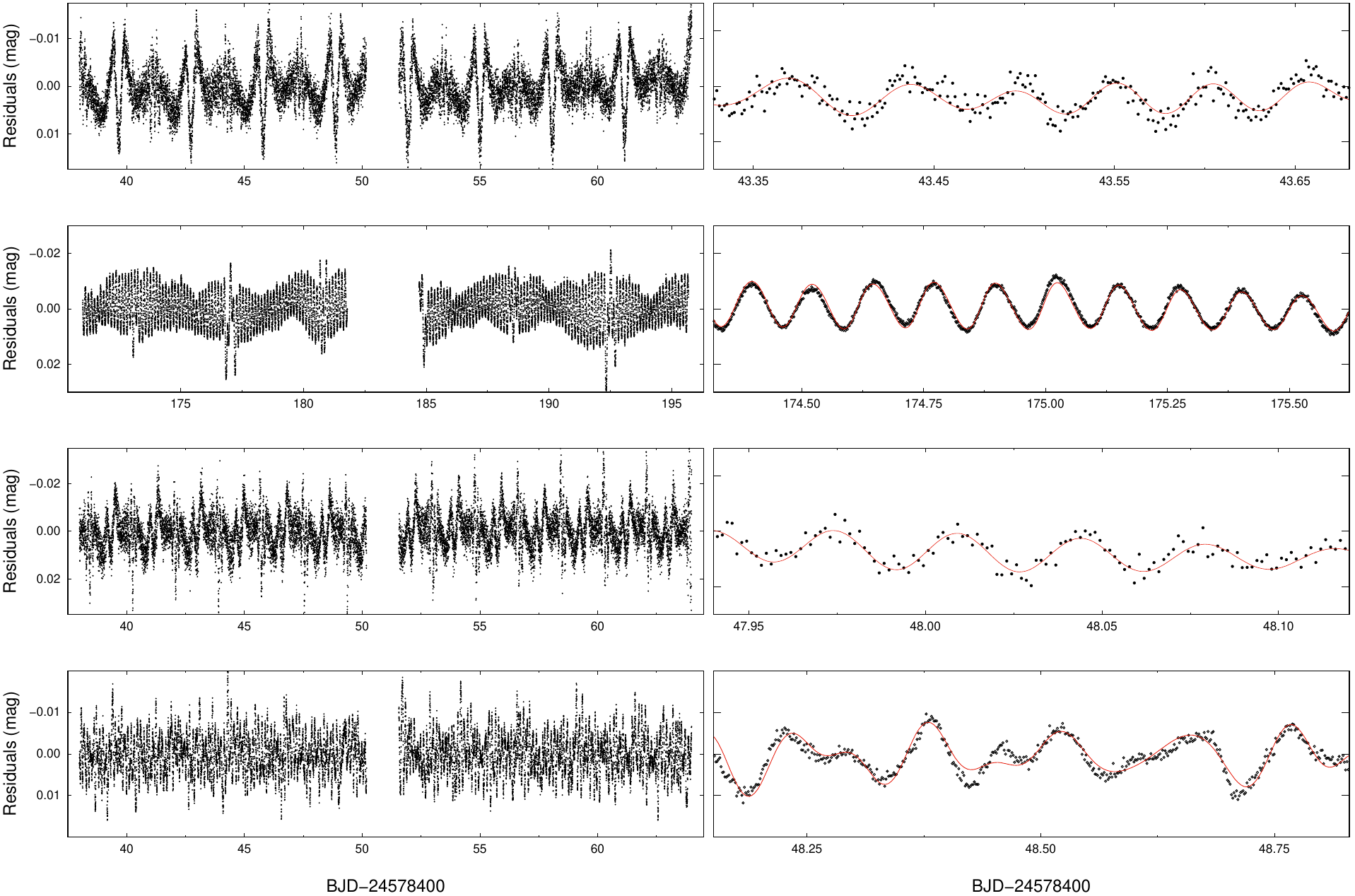}
\caption{Left panel: The residual light curves after removal of the binary models for CPD-30~740, HD~97329, V1637~Ori, and TYC~683-640-1 from top to bottom. Right panel: Agreement between Fourier fits (lines, red in colored version) and residual data in selected time intervals.}
\label{lcresfit}
\end{figure*}

After completing the binary modeling of the systems, we subtracted binary properties from the light curves to investigate the suspected pulsational behaviors affect the light curves. The residual data points are drawn versus time in the left panel of Fig.~\ref{lcresfit}. Following the derivation of the residuals, the frequency analyses applied to residual data of out-of-eclipse phases between 0 and 100~d$^{-1}$ by using {\tt PERIOD04} software \citep{len05} which employs Fourier analysis to the input data, extracts the main frequencies, and calculates the amplitude spectra. We prewhitened the derived frequency in every step and continue this process until reaching the signal-to-noise ratio of 4.0, which is the critical limit of the software. The calculated independent frequencies are listed in the following subsections, while the combination frequencies derived based on the Rayleigh criterion for individual data are tabulated in the Appendix. High-quality photometric data of TESS allows us to achieve very low significance levels. It is also worth pointing out that the data was taken in a long time duration with limited time gaps, and it is a primal advantage in terms of increasing the resolution of the solution in the frequency analysis process.

\subsection{CPD-30~740}\label{fr:cpd}
\begin{figure*}
\centering
\includegraphics[width=\textwidth]{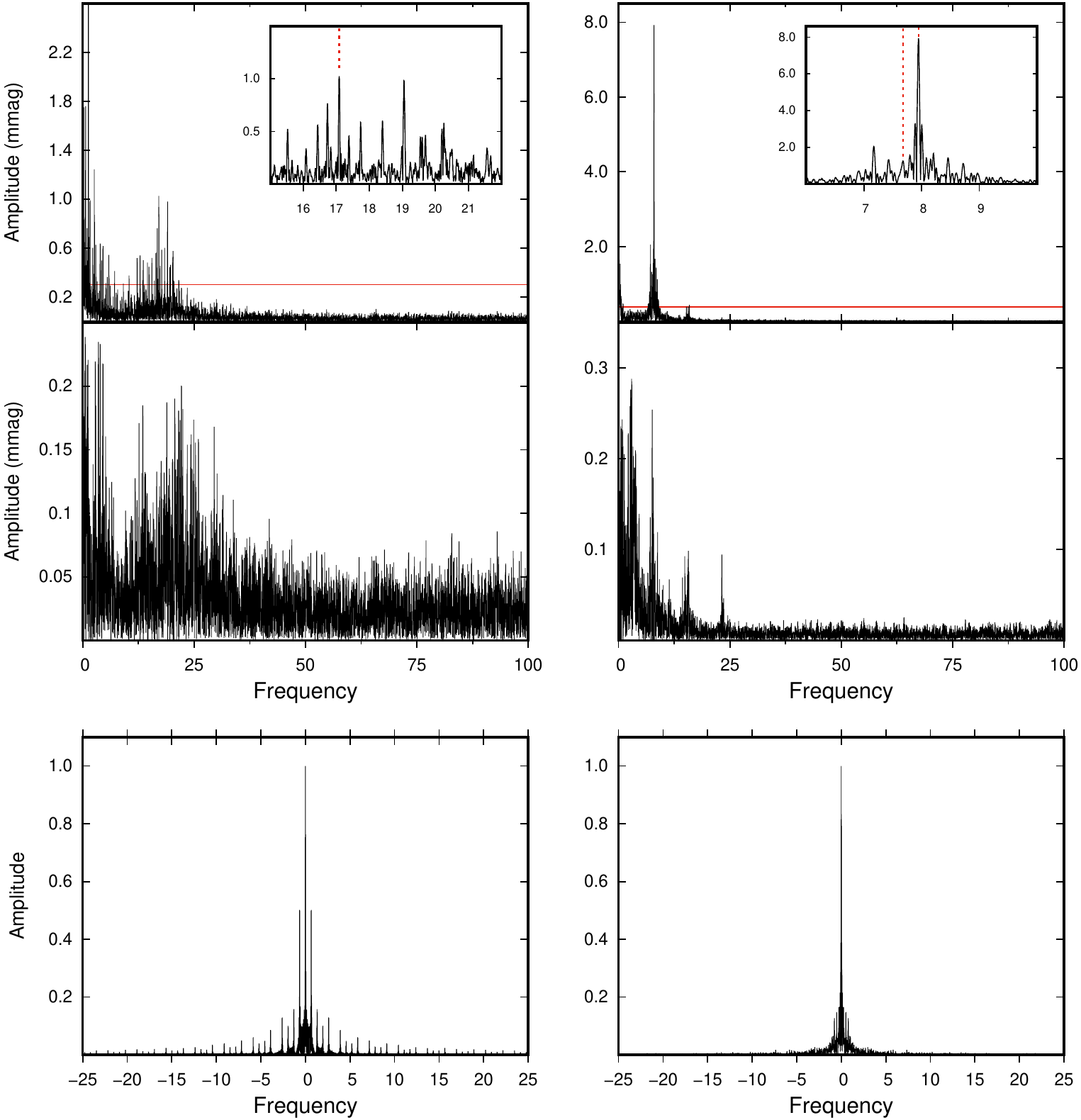}
\caption{Left panel, from top to bottom: Amplitude spectra before and after prewhitened all calculated frequencies and spectral window for {CPD-30~740}. Right panel: Same as left panel, but for HD~97329. The dashed lines (red in colored version) inlet represent the peaks with the genuine frequencies. Horizontal solid lines (red) refer to the significance limits for the data}
\label{lcfreq1}
\end{figure*}

We derive the residual light curve of the system by subtracting the observational data from the binary model with semi-detached configuration. A total of 10615 data points - after excluding minimum phases - were analyzed. The frequency analysis resulted in two genuine and 26 probable combination frequencies having signal-to-noise ratios are larger than 4.0. The genuine frequencies are tabulated in Table~\ref{tfr}. The analysis resulted in the frequency values smaller than 21~d$^{-1}$. The genuine frequency, $f_{4}$ is well inside the group of $\delta$~Sct type pulsators as it is larger than 5~d$^{-1}$ \citep{gri10}. The frequency smaller than 5~d$^{-1}$, $f_{3}$ of the genuine frequencies, are possibly rise from the effects of the binary removal process. It is, on the other hand, can be considered as the frequency of a $\gamma$~Dor type pulsation following the criteria of \cite{gri10} ($<$~5~d$^{-1}$). However, according to our results, the genuine frequencies do not fulfill the condition of being a hybrid pulsator remarked by \cite{uyt11}; two independent frequencies in both regimes must be observed. Eventually, the temperature of the primary component falls to the interval between A-F spectral types, namely the temperature range of the $\delta$~Sct stars. The results of the analysis also indicated that the independent frequency, $f_{4}$, is within the frequency range of $\delta$~Sct regime, 5-80~d$^{-1}$ \citep{bre00}. Additionally, we calculated the $Q$, the pulsation constants, to test the $\delta$~Sct-type pulsation of the component using the physical parameters in Table~\ref{tabs} and basic relation of pulsation and density:

\begin{align}
    Q=P_{p}\sqrt{\left( \frac{\rho}{\rho_{\odot}}\right)}
\end{align}
where $P_{p}$ is the period of pulsation and $\rho$ is the mean density of the pulsating component. The constant for the frequency $f_{4}$ 0.0161~d which is inside the range for typical $\delta$~Sct-type interval \citep[i.e. 0.015~$<Q<~$0.035;][]{bre00}. Thus far, it can be reported that CPD-30~740 is an eclipsing binary system with a pulsating primary component. Fig.~\ref{lcfreq1} shows the amplitude spectra and the spectral window of the analysis, while the agreement between the Fourier fit and the observations for the selected time interval is shown in Fig.~\ref{lcresfit}.

\begin{table}
\centering
\caption{Genuine frequencies derived during the frequency analyses. Parameters $f$, $A$, $\phi$, and SNR stand for frequency, amplitude, phase, and signal-to-noise ratio, respectively. The least-square uncertainties are given in the last digits. See Appendix for the list of possible combination frequencies derived by the analyses}
\label{tfr}
\begin{tabular}{lcccc}
\hline
& $f$ (d$^{-1}$) & $A$ (mmag) & $\phi$  &  SNR \\
\hline
\multicolumn{5}{c}{\underline{CPD-30~740}} \\
$f_{3} $&0.0236(6)&0.92(2)&0.173(4)&10\\
$f_{4} $&17.0888(5)&0.95(2)&0.155(4)&19\\
\hline
\multicolumn{5}{c}{\underline{HD~97329}} \\
$f_{1}$&7.94507(3)&8.26(1)&0.8211(2)&152\\
$f_{2}$&7.68159(12)&2.26(1)&0.9670(8)&44\\
\hline
\multicolumn{5}{c}{\underline{V1637~Ori}} \\
$f_{3}$&25.4863(4)&2.40(4)&0.372(3)&33\\
$f_{5}$&27.9692(6)&1.48(4)&0.056(5)&17\\
\hline
\multicolumn{5}{c}{\underline{TYC~683-640-1}} \\
$f_{1}$ &7.5628(2)&4.32(3)&0.876(1)&30\\
$f_{2}$ &11.1421(5)&1.43(3)&0.519(3)&8\\
$f_{3}$ &13.3113(6)&1.10(3)&0.867(4)&7\\
$f_{4}$ &10.3848(5)&1.43(3)&0.762(3)&9\\
$f_{5}$ &13.7629(6)&1.18(3)&0.946(4)&7\\
\hline
\end{tabular}
\end{table}

\subsection{HD~97329}
The residuals obtained by subtraction of the system's binary model cover 13403 data points. Our results show that there are 18 frequencies smaller than 16~d$^{-1}$ whose signal-to-noise ratios are higher than 4.0. Among them, 2 are genuine and the rest are potential combination frequencies. The genuine frequencies are listed in Table~\ref{tfr}. The results of the Fourier analysis are given in Fig.~\ref{lcfreq1} and the agreement of the final fit is plotted in Fig.~\ref{lcresfit}. The temperature of the hotter component and the derived genuine frequencies, 7.95 and 7.69~d$^{-1}$, which are larger than 5~d$^{-1}$, support the idea that the component is a $\delta$~Sct type pulsator. However, pulsation constants, $Q$ (Eq.~1), of genuine frequencies, 0.0597~d and 0.0617~d, are quite large to be within the interval for $\delta$~Sct stars given by \cite{bre00}. The genuine frequencies also do not satisfy the $\gamma$~Dor criteria, $<5~d^{-1}$. It can be remarked that the oscillation characteristics of the primary must be investigated in detail with advanced observational methods.

\subsection{V1637~Ori}
\begin{figure*}
\centering
\includegraphics[width=\textwidth]{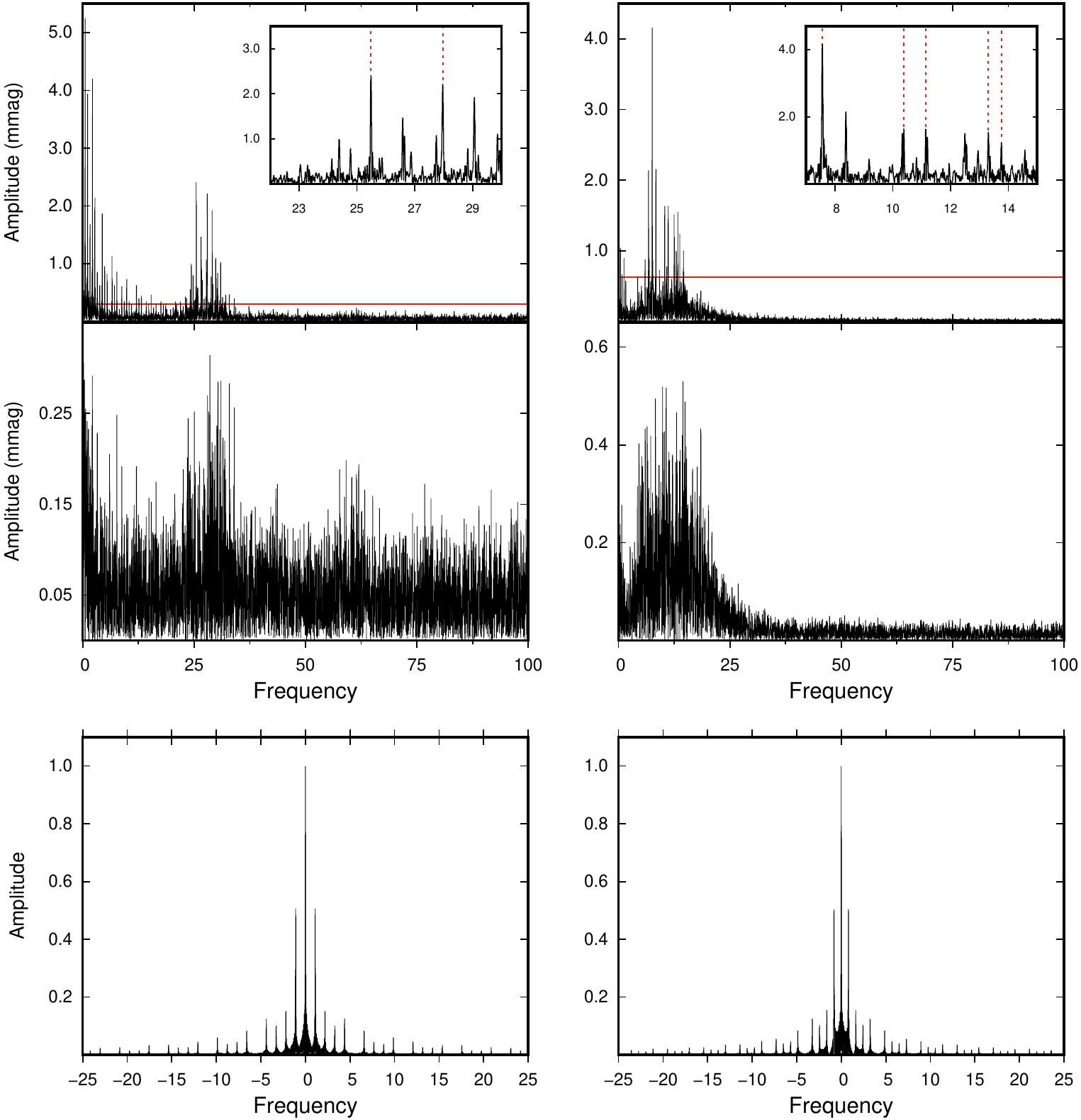}
\caption{Same as Fig.~\ref{lcfreq1}, but for V1637~Ori (left) and TYC~683-640-1 (right)}
\label{lcfreq2}
\end{figure*}

The number of data points obtained after removing the binary model is 10651. Our frequency analysis to residual data resulted in two genuine (25.4863 and 27.9692~d$^{-1}$), 25 combination frequencies with signal-to-noise ratio larger than 4.0. The maximum frequency value reached was 32~d$^{-1}$ during the analysis. The independent frequencies are listed in Table~\ref{tfr} and related figures plotted in Fig.~\ref{lcfreq2} while the Fourier fit plot with the residual data in Fig.~\ref{lcresfit}. The temperature of the primary component and the derived frequencies via analyses ($>$~5~d$^{-1}$) directed to us to infer that the primary component is a pulsating star of $\delta$~Sct type. Based on the absolute parameters in Table~\ref{tabs} and Eq.~1, the pulsation constants of the genuine frequencies were found to be 0.0172~d and 0.0157~d. These values endorse the membership of this type of pulsational class, although the latter one is very close to the lower limit. Therefore, the primary component of the system is strongly considered as a $\delta$~Sct star in the light of the above findings.

\subsection{TYC~683-640-1}
TYC~683-640-1 is the system having five independent (Table~~\ref{tfr}), 13 combination frequencies derived by the frequency analysis which was applied to 10649 residual data points of maximum phases. The temperature of the primary component is inside the $\delta$~Sct type interval and genuine frequencies are larger than 5~d$^{-1}$. Furthermore, pulsation constants, calculated by Eq.~1, can be considered within the  $\delta$~Sct range; 0.0259, 0.0175, 0.0147, 0.0188 and 0.0142 for $f_{1}$, $f_{2}$, $f_{3}$ and $f_{4}$ and $f_{5}$, respectively. The amplitude spectra and the spectral window are represented in Fig.~\ref{lcfreq2}. Fourier fit is represented with the data in Fig.~\ref{lcresfit}.

\begin{figure}
\centering
\includegraphics{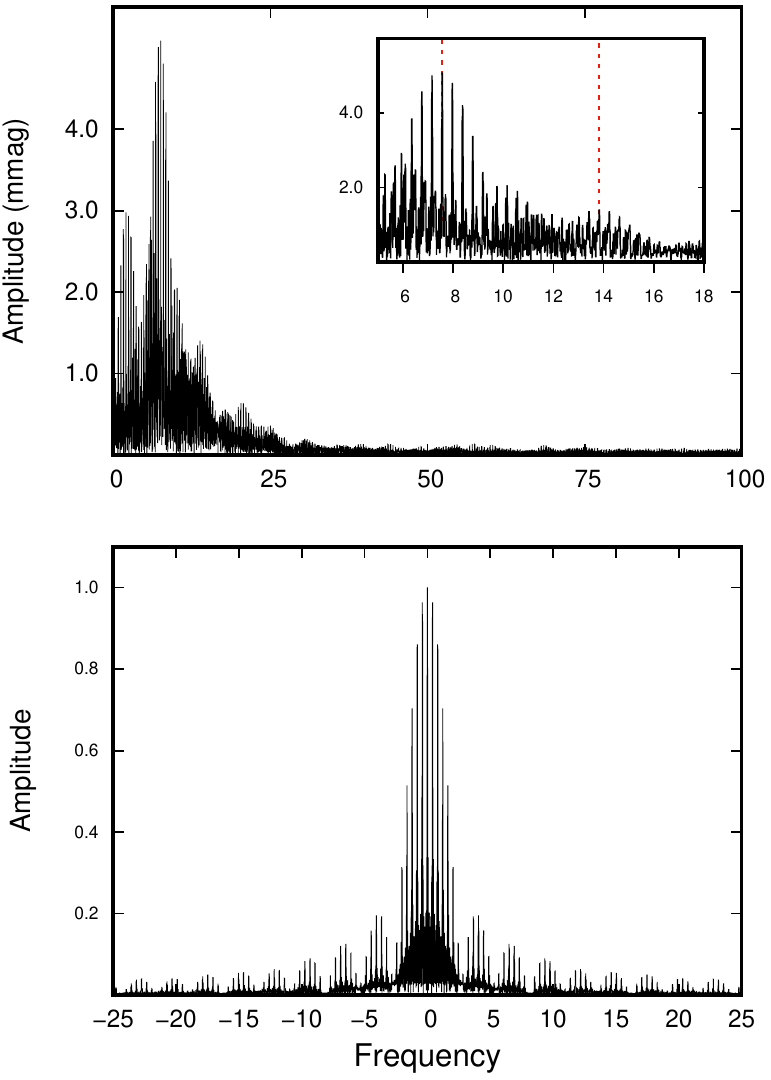}
\caption{Amplitude spectra and spectral window for the secondary minimum phases of the light curve of TYC~683-640-1 where only the primary component is observed. The dashed lines (red in colored version) inlet represent the peaks with the genuine frequencies}
\label{frmin2}
\end{figure}

Since the light curve of the system shows total eclipse, a frequency analysis can be applied to data at phases of secondary minimum where only the primary component is observed. Thuswise, the dominance of the component to the pulsational behavior can be tested. Therefore, we analyzed the data covering the secondary minimum phases and found out that the amplitude (4.76 mmag) of the first genuine frequency (7.5680~d$^{-1}$) which is as large as that of 7.5628~d$^{-1}$, found by analyzing the data of maximum phases (Table~\ref{tfr}). We also derived another independent frequency, 13.8275~d$^{-1}$, having almost the same amplitude with $f_3$ in Table~\ref{tfr}. The amplitude spectrum and the spectral window for the frequency analysis of the secondary minimum phases are given in Fig~\ref{frmin2}. Herewith, we can conclude that the system is a detached binary with a $\delta$~Sct type primary component.

\section{Conclusion}

A detailed investigation of the binary and pulsational properties of four binary systems is presented. We applied light curve and frequency analyses and tried to reveal the possible pulsational behavior of one of the components for the first time in the literature. We also calculated the absolute parameters for the systems based on our results by using {\tt{ABsParEB}} software \citep{lia15a}. The masses of the primary components of our semi-detached targets, CPD-30~740 and V1637~Ori, were estimated based on their log~$g$ and effective temperature values among 204930 binary star models that we produced by using The Binary Star Evolution code \citep[{\tt{BSE}},][]{hur02,hur13} with solar abundance assumption. The models obtained by running the code with the various initial parameter values of eccentricity between 0 and 0.5, orbital period of the system between 0.5~d and 7~d, mass of the primary component, $M_1$, between 0.5 M$_{\odot}$ and 5.0 M$_{\odot}$, and mass of the secondary component between 0.1 M$_{\odot}$ and $M_1$. The mass for the primary component of TYC~683-640-1 are adopted from the stellar tracks of \cite{ber09} according to its log~$g$ and effective temperature values by assuming that the star is in solar abundance. The masses of HD~97329 were, on the other hand, calculated by considering the amplitudes, $K_1$=88$\pm$3~kms$^{-1}$ and $K_2$=79$\pm2$~kms$^{-1}$, which are derived from the synthetic radial velocities curve and the orbital inclination value. The absolute parameters for the systems are given in Table~\ref{tabs}.

\begin{table*}
\centering
\caption{Absolute parameters of the systems. The standard errors are given in parentheses for the last digits. P and S denote the primary and secondary components, respectively}
\label{tabs}
\begin{tabular}{lcccccccc}
\hline
Parameter                &\multicolumn{2}{c}{CPD-30~740} &\multicolumn{2}{c}{HD~97329} & \multicolumn{2}{c}{V1637~Ori}  &  \multicolumn{2}{c}{TYC~683-640-1}  \\
\hline 
& \multicolumn{1}{c}{\underline{P}}&\multicolumn{1}{c}{\underline{S}}&\multicolumn{1}{c}{\underline{P}}&\multicolumn{1}{c}{\underline{S}} &\multicolumn{1}{c}{\underline{P}}&\multicolumn{1}{c}{\underline{S}} &\multicolumn{1}{c}{\underline{P}}&\multicolumn{1}{c}{\underline{S}} \\
M (M$_{\odot}$) & 1.6 &  0.900(2)    & 1.8(1)& 2.0(3)  & 2.1& 1.166(4)          & 1.8& 0.738(2)\\
R (R$_{\odot}$) & 2.77(6)&3.69(7)    & 2.5(9)& 2.8(8)  & 2.22(9)& 3.12(2)        & 3.61(4)& 1.1(3)\\
L (L$_{\odot}$) & 16.5(8) &  6.7(3)  & 12(9)& 7(4)     & 23(2)& 6.85(8)           &42.1(8) & 2(1)\\
$a $ (R$_{\odot}$) & \multicolumn{2}{c}{12.408(7)} & \multicolumn{2}{c}{25.7(5)} & \multicolumn{2}{c}{9.53(1)} & \multicolumn{2}{c}{10.711(7)}\\
\hline
\end{tabular}
\end{table*}

\begin{figure*}
\centering
\includegraphics[width=\textwidth]{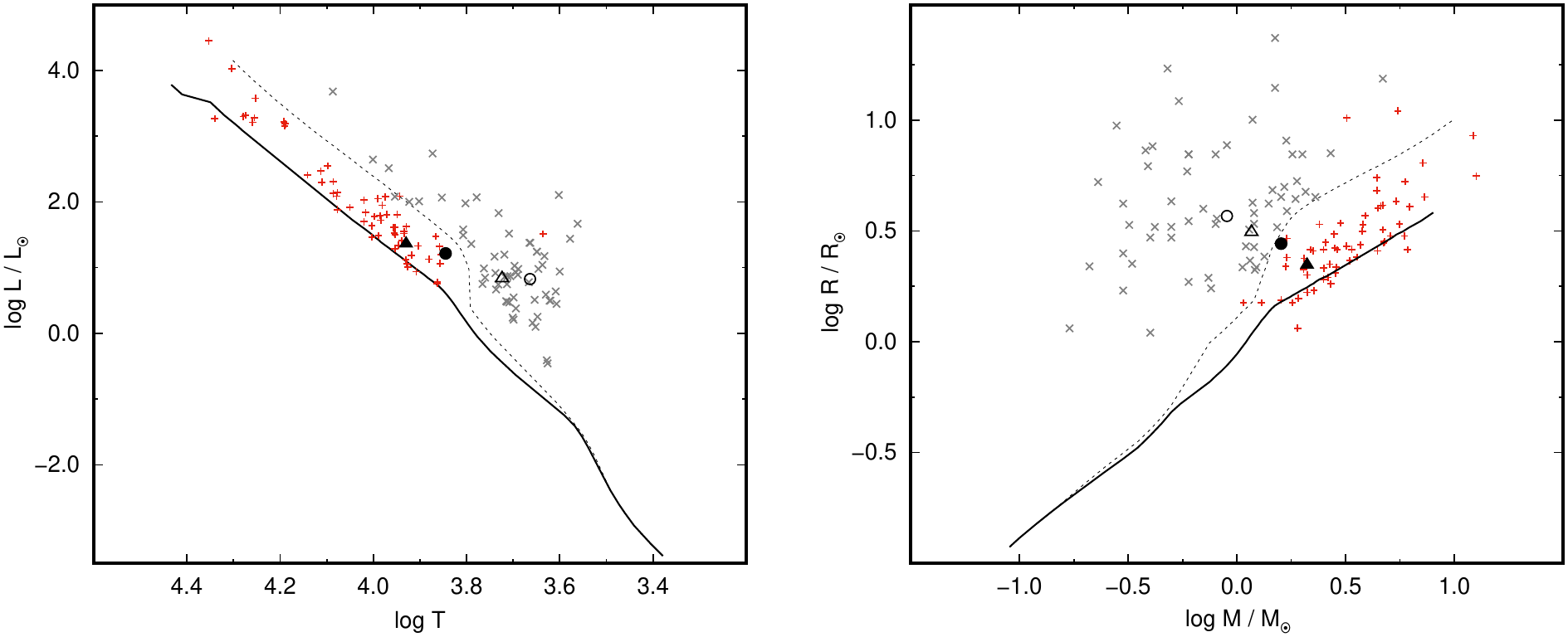}
\caption{Location of the components of  CPD-30~740 and V1637~Ori on the Hertzsprung-Russell diagram (left) and mass-radius plane (right). Plus and crosses (pale colors; grey and tan in colored version) refer to the primary and secondary components of known Algols \citep{iba06}. Filled signs denote the primary components of the systems, where the open signs assign the secondaries. Circles and triangles refer to the components of CPD-30~740 and V1637~Ori. The data for ZAMS (thick solid line) and TAMS (dashed line) are taken from \cite{bre12} with Z=0.01 and Y=0.267 abundance}
\label{fhrmralg}
\end{figure*}

\begin{figure*}
\centering
\includegraphics[width=\textwidth]{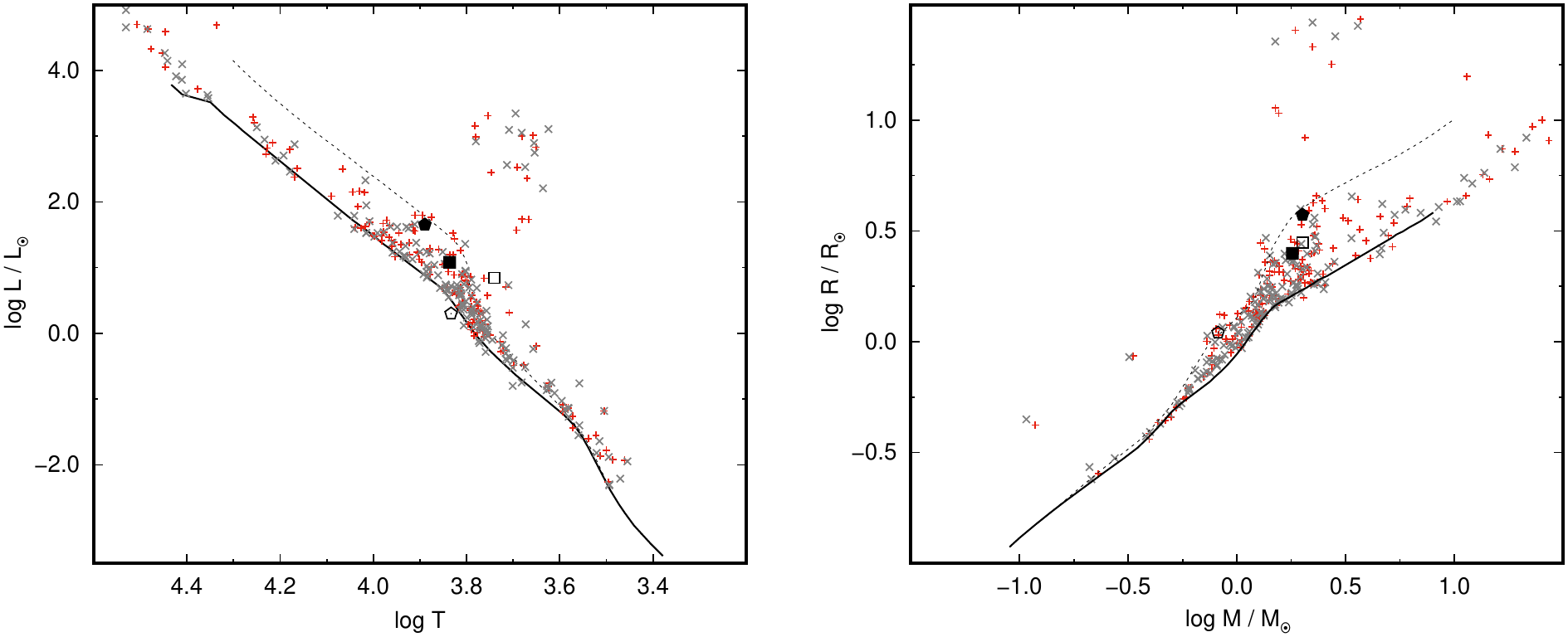}
\caption{Same as Fig.~\ref{fhrmralg}, but for HD~97329 and TYC~683-640-1. Plus and crosses (pale colors; grey and tan in colored version) refer to the primary and secondary components of known detached systems cataloged by \cite{sou15}. Filled signs denote the primary components of the systems, where the open signs assign the secondaries. Squares and pentagons refer to the components of HD~97329 and TYC~683-640-1}
\label{fhrmrdet}
\end{figure*}

We compare two of our targets, CPD-30~740 and V1637~Ori, to well-known Algol type binaries given by \cite{iba06} in the mass-radius plane and the Hertzsprung-Russell diagram (Fig.~\ref{fhrmralg}). The components are in good agreement with the other Algols. The two other systems, HD~97329 and TYC~683-640-1, are also compared to 162 detached systems of \cite{sou15} in Fig.~\ref{fhrmrdet}. The secondary component of HD~97329 placed slightly luminous location according to its temperature on the Hertzsprung-Russell diagram, while the secondary component of TYC~683-640-1 is located mildly underluminous position according to its effective temperature.

\begin{figure*}
\centering
\includegraphics[width=\textwidth]{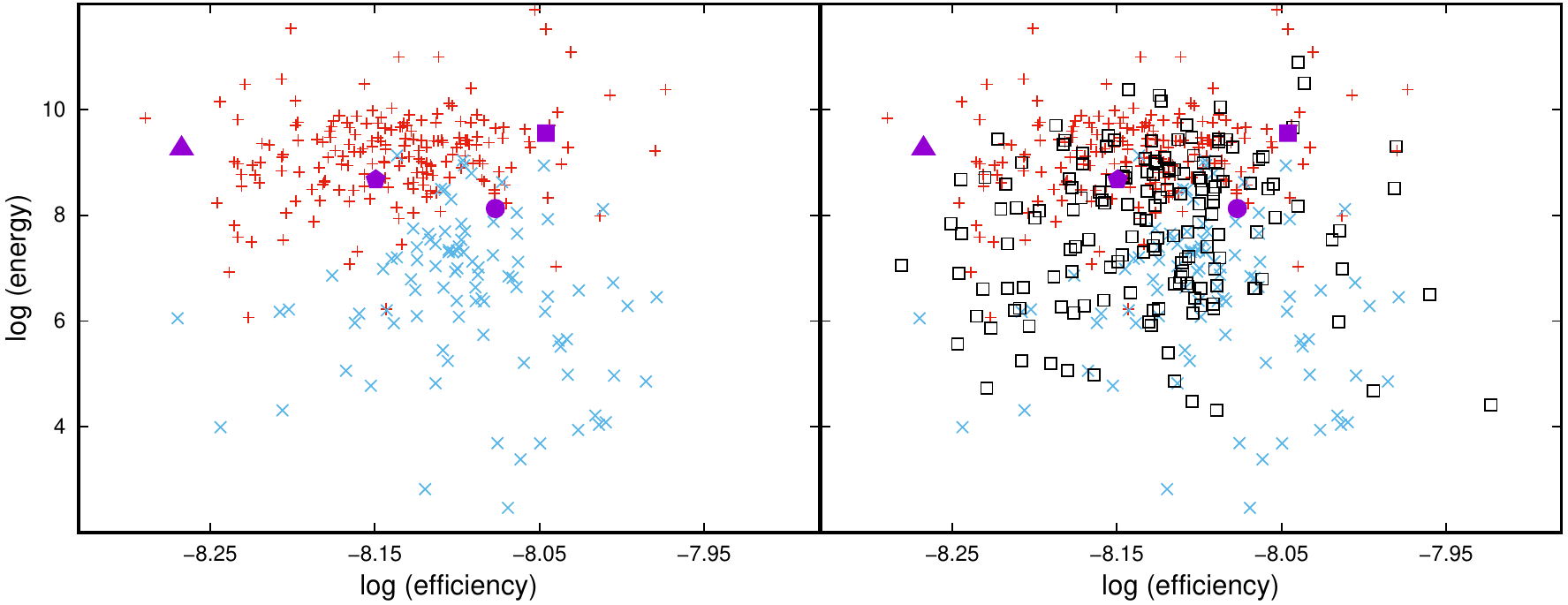}
\caption{Location of the pulsating components on log(efficiency)-log(energy) diagrams without (left) and with (right) hybrid $Kepler$ pulsators. Plus (red in colored version), cross (blue), and open squares (black) refer $\delta$~Sct, $\gamma$~Dor, and hybrid stars, respectively. Full (purple) signs denote the primary components of CPD-30~740 (circle), HD~97329 (square), V1637~Ori (triangle) and TYC~683-640-1 (pentagon). The data for $\delta$~Sct, $\gamma$~Dor, and hybrid stars are taken from \cite{uyt11}}
\label{lcee}
\end{figure*}

Although we inferred the types of oscillations for the possible pulsating components in related subsections, we also located the components on the efficiency-energy diagram for $Kepler$ pulsators given by \cite{uyt11} to look for any other clue that is expected to give us information about pulsation classes for the components among the three types of pulsators, $\delta$~Sct, $\gamma$~Dor, and hybrid. For doing this, we converted the fluxes to ppm and calculated the efficiency and energy parameters following the formulation given by \cite{uyt11}; $efficiency\equiv (T_{eff} ^{3} \log g)^{-2/3}$ and $energy\equiv (A_{max}\zeta_{max})^2$, where $A_{max}$ is the highest amplitude mode (in ppm) having the frequency value of $\zeta_{max}$ and the others denote the standard quantities. As seen from the left panel of Fig~\ref{lcee}, we may remark that the primary components of HD~97329, V1637~Ori and TYC~683-640-1 fall inside the $\delta$~Sct distribution although the efficiency value of the primary of V1637~Ori is lower than many stars of the same type. Despite the fact that it is a $\delta$~Sct type pulsator (See \ref{fr:cpd}), the diagram brings the pulsation class of the primary component of CPD-30~740 into question since the star is nested on the boundary between $\delta$~Sct and $\gamma$~Dor types. Additionally, because of the wide distribution of hybrid class (right panel of Fig.~\ref{lcee}), it seems not appropriate to judge the hybrid pulsator candidacy of the targets by the diagram.

As a closing remark, we note that our targets are eclipsing binary systems with pulsating components. Further, CPD-30~740 and V1637~Ori are found to be the members of the group of oEAs \citep{mkr02, mkr04} since the binary modelling corresponds that the systems are classical semi-detached Algol type binaries. TYC~683-640-1 is a detached binary system having a $\delta$~Sct primary. Our analyses show that the primary component of HD~97329 shows certain pulsation behavior, however, the type of its oscillation could not be determined with the methods used. Considering the importance of the physical parameters in determination of the binary and pulsation characteristics, observational method like high-quality spectroscopy is extremely needed to test our results and relieve the properties of the systems more precisely.

        %
        %

\section*{Acknowledgement}
The authors would like to thank Dr. David Mkrtichian for his valuable comments and suggestions. CU acknowledges the research grant No.~R00125 administered by the office of the DVC Research, Development and Innovation at Botswana International University of Science and Technology (BIUST). Southern Cross Astronomy Research Group would like to thank BIUST for the support from grant No.~R00125. This paper includes data collected with the TESS mission, obtained from the MAST data archive at the Space Telescope Science Institute (STScI). Funding for the TESS mission is provided by the NASA Explorer Program. STScI is operated by the Association of Universities for Research in Astronomy, Inc., under NASA contract NAS 5–26555. This research has made use of the SIMBAD database, operated at CDS, Strasbourg, France. Some/all of the data presented in this paper were obtained from the Mikulski Archive for Space Telescopes (MAST). STScI is operated by the Association of Universities for Research in Astronomy, Inc., under NASA contract NAS5-26555. Support for MAST for non-HST data is provided by the NASA Office of Space Science via grant NNX13AC07G and by other grants and contracts. This research has made use of the VizieR catalogue access tool, CDS, Strasbourg, France. This work has made use of data from the European Space Agency (ESA) mission {\it Gaia} (\url{https://www.cosmos.esa.int/gaia}), processed by the {\it Gaia} Data Processing and Analysis Consortium (DPAC, \url{https://www.cosmos.esa.int/web/gaia/dpac/consortium}). Funding for the DPAC has been provided by national institutions, in particular the institutions participating in the {\it Gaia} Multilateral Agreement. This work has made use of the NASA/IPAC Infrared Science Archive, which is funded by the National Aeronautics and Space Administration and operated by the California Institute of Technology.

\begin{appendix}
\section{List of combined frequencies}

\setcounter{table}{0}
\renewcommand\thetable{\Alph{section}.\arabic{table}}

Table~\ref{ta1} shows the possible combination frequencies of the primary components of the systems as a continuation of Table~\ref{tfr}. The combination relations are denoted in the first column.

\begin{center}
\small
\begin{longtable}{lcccc}
\caption{Combination frequencies derived from the frequency analysis of the systems. The standard errors are given in parentheses for the last digits.} \label{tab:long} \label{ta1}
\\

\hline 
& $f$ (d$^{-1}$) & $A$ (mmag) & $\phi$  &  SNR \\

\hline 
\endfirsthead

\multicolumn{3}{c}%
{{\tablename\ \thetable{} -- continued from previous page}} \\
\hline 
& $f$ (d$^{-1}$) & $A$ (mmag) & $\phi$  &  SNR \\ \hline 
\endhead

\hline \multicolumn{4}{r}{{Continued on next page}} \\
\endfoot

\hline \hline
\endlastfoot
\multicolumn{5}{c}{\underline{CPD-30~740}} \\
$f_{1} \approx 4f_{orb}$&1.3039(2)&3.23(2)&0.993(1)&50\\
$f_{2} \approx f_{orb}$&0.3299(3)&2.01(2)&0.641(2)&24\\
$f_{5} \approx f_{1}+2f_{2}$&1.9579(4)&1.36(2)&0.521(3)&24\\
$f_{6} \approx f_{4}+f_{5}$&19.0506(5)&1.04(2)&0.476(4)&15\\
$f_{7} \approx 5f_{orb}$&1.6299(5)&1.09(2)&0.693(3)&18\\
$f_{8} \approx f_{3}$&0.0648(7)&0.73(2)&0.901(5)&8\\
$f_{9} \approx f_{4}-f_{2}$&16.7333(8)&0.69(2)&0.375(6)&14\\
$f_{10} \approx f_{1}+f_{6}-f_{8}$&20.2583(10)&0.52(2)&0.773(7)&8\\
$f_{11} \approx f_{9}-2f_{5}$&12.8765(9)&0.58(2)&0.207(7)&11\\
$f_{12} \approx f_{11}+2f_{1}$&15.5276(7)&0.69(2)&0.871(6)&14\\
$f_{13} \approx f_{11}+f_{7}$&14.5280(12)&0.45(2)&0.511(9)&9\\
$f_{14} \approx f_{12}-f_{5}$&13.5677(14)&0.38(2)&0.586(10)&7\\
$f_{15} \approx f_{6}$&19.0958(15)&0.35(2)&0.927(11)&5\\
$f_{16} \approx f_{14}+f_{4}-f_{10}$&10.3845(14)&0.37(2)&0.209(10)&9\\
$f_{17} \approx f_{11}-2f_{1}$&10.2509(16)&0.33(2)&0.869(11)&8\\
$f_{18} \approx f_{10}$&20.3035(16)&0.32(2)&0.486(12)&5\\
$f_{19} \approx 2f_{17}$&20.4999(14)&0.37(2)&0.698(10)&5\\
$f_{20} \approx f_{19}+f_{2}-f_{1}$&19.5435(15)&0.34(2)&0.676(11)&5\\
$f_{21} \approx f_{13}+f_{7}$&16.1874(13)&0.41(2)&0.813(9)&9\\
$f_{22} \approx f_{1}+f_{10}$&21.5642(16)&0.32(2)&0.407(12)&4\\
$f_{23} \approx 10f_{orb}$&3.2618(18)&0.29(2)&0.766(13)&5\\
$f_{24} \approx f_{19}$&20.4527(18)&0.28(2)&0.268(14)&4\\
$f_{25} \approx 2f_{orb}$&0.7050(13)&0.39(2)&0.309(10)&5\\
$f_{26} \approx f_{2}+f_{5}$&2.2819(11)&0.45(2)&0.082(8)&9\\
$f_{27} \approx f_{11}-f_{1}$&11.5863(21)&0.25(2)&0.525(15)&6\\
$f_{28} \approx f_{1}-f_{25}$&0.5774(14)&0.36(2)&0.543(11)&4\\

\hline
\multicolumn{5}{c}{\underline{HD~97329}} \\
$f_{3}\approx f_{orb}$&0.13377(18)&1.50(1)&0.4238(13)&15\\
$f_{4}\approx 3f_{3}$&0.38712(22)&1.21(1)&0.4588(16)&13\\
$f_{5}\approx f_{1}-f_{3}$&7.83765(26)&1.02(1)&0.7997(19)&19\\
$f_{6}\approx f_{1}-2f_{4}$&7.18097(34)&0.79(1)&0.2898(24)&16\\
$f_{7}\approx 2f_{3}$&0.28578(39)&0.69(1)&0.9111(28)&7\\
$f_{8}\approx f_{2}$&7.71402(59)&0.45(1)&0.8235(42)&9\\
$f_{9}\approx f_{5}-f_{6}$&0.65668(63)&0.42(1)&0.1541(45)&4\\
$f_{10}\approx 2f_{1}$&15.89218(57)&0.47(1)&0.2701(41)&17\\
$f_{11}\approx f_{4}+f_{9}$&1.02151(59)&0.45(1)&0.8244(42)&5\\
$f_{12}\approx 2f_{5}$&15.63072(63)&0.43(1)&0.5582(45)&14\\
$f_{13}\approx f_{2}-f_{6}$&0.44792(42)&0.64(1)&0.6589(30)&7\\
$f_{14}\approx f_{12}-f_{4}$&15.24157(70)&0.39(1)&0.3584(50)&13\\
$f_{15}\approx 2f_{3}$&0.21889(63)&0.43(1)&0.1437(45)&4\\
$f_{16}\approx f_{1}+f_{3}$&8.07276(79)&0.34(1)&0.2399(56)&6\\
$f_{17}\approx 2f_{7}$&0.52494(69)&0.39(1)&0.7104(49)&4\\
$f_{18}\approx f_{1}-f_{13}$&7.44851(82)&0.33(1)&0.2339(58)&6\\
\hline
\multicolumn{5}{c}{\underline{V1637~Ori}} \\
$f_{1} \approx f_{orb}$&0.5489(3)&3.31(4)&0.385(2)&26\\
$f_{2} \approx 4f_{orb}$&2.1958(1)&6.56(4)&0.628(1)&70\\
$f_{4} \approx 6f_{orb}$&3.2917(2)&4.77(4)&0.155(1)&68\\
$f_{6} \approx 5f_{orb}$&2.7447(4)&2.43(4)&0.594(3)&30\\
$f_{7} \approx 8f_{orb}$&4.3896(5)&1.96(4)&0.865(4)&32\\
$f_{8} \approx f_{2}+f_{3}+f_{4}$&30.9719(12)&0.79(4)&0.149(9)&9\\
$f_{9} \approx 2f_{5}-f_{3}-f_{6}$&27.7461(6)&1.48(4)&0.301(5)&18\\
$f_{10} \approx f_{4}+f_{9}-f_{2}$&28.8382(8)&1.25(4)&0.546(6)&15\\
$f_{11} \approx f_{4}+f_{5}+f_{2}$&29.0651(8)&1.19(4)&0.091(6)&14\\
$f_{12} \approx f_{5}+f_{9}-f_{8}$&24.7802(13)&0.74(4)&0.060(9)&10\\
$f_{13} \approx f_{12}+f_{2}+f_{4}$&30.3008(13)&0.74(4)&0.153(10)&8\\
$f_{14} \approx f_{7}-2f_{2}$&0.0291(15)&0.62(4)&0.348(11)&5\\
$f_{15} \approx f_{5}$&27.9362(13)&0.71(4)&0.665(10)&8\\
$f_{16} \approx f_{5}-f_{2}-f_{6}$&23.0422(15)&0.61(4)&0.053(11)&9\\
$f_{17} \approx 9f_{orb}$&29.8566(16)&0.60(4)&0.318(12)&7\\
$f_{18} \approx 7f_{orb}$&3.8446(10)&0.91(4)&0.530(8)&13\\
$f_{19} \approx f_{orb}$&0.5218(18)&0.51(4)&0.670(14)&4\\
$f_{20} \approx f_{11}+f_{6}$&31.8138(21)&0.45(4)&0.297(16)&5\\
$f_{21} \approx 12f_{orb}$&6.5874(23)&0.40(4)&0.957(17)&7\\
$f_{22} \approx f_{14}+f_{4}+f_{9}$&31.1232(23)&0.42(4)&0.923(17)&5\\
$f_{23} \approx f_{22}-f_{7}$&26.7374(25)&0.38(4)&0.204(18)&5\\
$f_{24} \approx f_{1}+f_{23}$&27.2689(27)&0.34(4)&0.457(20)&4\\
$f_{25} \approx f_{13}-f_{18}$&26.4562(26)&0.36(4)&0.611(20)&5\\
$f_{26} \approx f_{25}-f_{4}$&23.2187(30)&0.32(4)&0.687(22)&4\\
$f_{27} \approx f_{1}+f_{5}$&28.5647(30)&0.32(4)&0.867(22)&4\\
\hline
\multicolumn{5}{c}{\underline{TYC~683-640-1}} \\
$f_{6} \approx f_{orb}$&0.4122(5)&1.38(3)&0.412(3)&13\\
$f_{7} \approx f_{3}+f_{4}-f_{2}$&12.5540(8)&0.81(3)&0.172(6)&5\\
$f_{8} \approx f_{1}+f_{4}-f_{2}$&6.7740(7)&0.89(3)&0.079(5)&6\\
$f_{9} \approx f_{8}$&6.8568(8)&0.83(3)&0.596(6)&5\\
$f_{10} \approx 2f_{orb}$&0.8125(7)&0.94(3)&0.952(5)&10\\
$f_{11} \approx f_{4}+f_{7}-f_{1}$&15.3859(11)&0.61(3)&0.039(8)&4\\
$f_{12} \approx f_{3}-f_{10}$&12.4594(10)&0.69(3)&0.827(7)&4\\
$f_{13} \approx f_{1}-f_{6}$&7.1723(11)&0.59(3)&0.640(8)&4\\
$f_{14} \approx f_{3}-f_{6}$&12.9386(9)&0.73(3)&0.772(7)&4\\
$f_{15} \approx f_{10}+f_{5}$&14.6148(12)&0.53(3)&0.223(9)&3\\
$f_{16} \approx f_{7}$&12.5047(9)&0.77(3)&0.126(6)&5\\
$f_{17} \approx f_{1}+f_{9}$&14.4768(3)&2.57(3)&0.058(2)&15\\
$f_{18} \approx f_{17}$&14.4748(3)&2.55(3)&0.386(2)&14\\
\end{longtable}
\end{center}
\end{appendix}

\twocolumn

\section*{Declarations}
\subsection*{Funding}
This work was funded by Botswana International University of Science and Technology (BIUST) (Grant number R00125).

\begin{sloppypar}
\bibliographystyle{spr-mp-nameyear-cnd}
\bibliography{Ulas_bib}

\end{sloppypar}

\end{document}